\PassOptionsToPackage{usenames,dvipsnames,svgnames,table}{xcolor}
\documentclass[sigconf,authorversion]{acmart}

\usepackage[utf8]{inputenc}
\usepackage{csquotes}
\usepackage{enumitem}
\usepackage{tabularx,multirow,multirow}
\usepackage{balance}
\usepackage[capitalise,nameinlink]{cleveref}

\setcopyright{acmlicensed}
\copyrightyear{2022} 
\acmYear{2022}
\acmConference[ITiCSE 2022] {Proceedings of the 27th ACM Conference on Innovation and Technology in Computer Science Education Vol 1}{July 8--13, 2022}{Dublin, Ireland.}
\acmBooktitle{Proceedings of the 27th ACM Conference on Innovation and Technology in Computer Science Education Vol 1 (ITiCSE 2022), July 8--13, 2022, Dublin, Ireland}
\acmPrice{15.00}
\acmDOI{10.1145/3502718.3524821}
\acmISBN{978-1-4503-9201-3/22/07}

\begin{document}

\fancyhead{}

\title[Experience with Remote Teaching of Embedded Systems]{Experience with Abrupt Transition to Remote Teaching\\of Embedded Systems}

\author{Jan Koniarik}
\affiliation{
  \institution{Masaryk University}
  \city{Brno}
  \country{Czech Republic}
}
\email{433337@mail.muni.cz}
\orcid{0000-0001-6824-6734}

\author{Daniel Dlhopolček}
\affiliation{
  \institution{Masaryk University}
  \city{Brno}
  \country{Czech Republic}
}
\email{xdlhopol@mail.muni.cz}

\author{Martin Ukrop}
\affiliation{
  \institution{Masaryk University}
  \city{Brno}
  \country{Czech Republic}
}
\email{mukrop@mail.muni.cz}
\orcid{0000-0001-8110-8926}

\begin{abstract}
Due to the pandemic of COVID-19, many university courses had to abruptly transform to enable remote teaching. Adjusting courses on embedded systems and micro-controllers was extra challenging since interaction with real hardware is their integral part. We start by comparing our experience with four basic alternatives of teaching embedded systems: 1) interacting with hardware at school, 2) having remote access to hardware, 3) lending hardware to students for at-home work and 4) virtualizing hardware. Afterward, we evaluate in detail our experience of the fast transition from traditional, offline at-school hardware programming course to using remote access to real hardware present in the lab. The somewhat unusual remote hardware access approach turned out to be a fully viable alternative for teaching embedded systems, enabling a relatively low-effort transition. Our setup is based on existing solutions and stable open technologies without the need for custom-developed applications that require high maintenance. We evaluate the experience of both the students and teachers and condense takeaways for future courses. The specific environment setup is available online as an inspiration for others.
\end{abstract}



\maketitle

\section{Introduction}
\label{sec:introduction}

Interest in blended learning (combination of online and face-to-face) was shown in surveys already at the beginning of the 21\textsuperscript{st} century~\cite{garrison2012blended}. Later on, the use of online learning dramatically increased worldwide with the approach of the COVID-19 pandemic in early 2020~\cite{online-education-covid}. However, the inclusion of information technology and online services into education can be done in multiple ways: While innovative use improves education, simple substitutive use can even decrease the quality of learning~\cite{garrison2012blended}. Due to the sudden nature of the change, Adedoyin and Soykan~\cite{emergency-remote-teaching} suggest calling online education at the beginning of the pandemic by the term \enquote{emergency remote teaching} to put it in contrast with the thoroughly prepared, efficient online teaching.

This paper reflects on our experience with a fast transformation of a university-level course on embedded hardware programming for emergency remote teaching. Since the university decided to have a fully remote semester only four weeks before the semester started, we needed a fast transition. With at-school hardware interaction no longer possible, we saw three basic setups to teach hardware systems: 1) distributing hardware to students for work at home, 2) using purely virtualized hardware, or 3) remotely accessing hardware at school.

Not having enough hardware to distribute to students individually and wanting to retain as many hardware interactions as possible (it is one of the few hardware-based courses at the faculty), we decided to set up remote access to hardware in the lab. Previous works using remote hardware access~\cite{astatke2012,kulich2013} were almost exclusively based on complex custom-made platforms developed in-house over the years. We aimed for a much simpler solution that could be set up quickly and would require low maintenance while allowing us to make as few changes to the course as possible (compared to the in-person teaching). We wanted to entirely avoid the development of new custom tools, reusing existing infrastructure, and stable, open software.

In summary, our paper attempts to look for answers to the following two research questions:
\begin{enumerate}[leftmargin=1.5em,itemsep=1em]
    \item \textbf{How do different setups of teaching embedded systems compare?}
    We try to compare and contrast the main features of four principally different approaches to hardware access based on our ten years-long pedagogical experience in the field.
    \item \textbf{Is education based on remote hardware access viable even without complex custom-made systems? What are its benefits and challenges?}
    We show that remote hardware access is an effective approach for teaching hardware, deployable in a matter of weeks using existing technologies. We describe our setup for others to adapt and summarize challenges encountered and lessons learned.
\end{enumerate}
After describing the related work in \cref{sec:related-work}, we look into the context of the course and compare possible high-level approaches to hardware access in \cref{sec:course-redesign}. The section ends with a description of our transformation. Reusing existing infrastructure and standard technology stack, we set up virtual machines in the lab with direct access to the hardware. Remote access is provided using university VPN with visual feedback via connected webcams.
In \cref{sec:evaluation}, we evaluate the transformation from the perspective of both students and teachers. Lessons learned are summarized in \cref{sec:lessons-learned}.

\section{Related Work}
\label{sec:related-work}

As stated by \citeauthor{chung2018} in 2018~\cite{chung2018}, previous research on how teaching online affects computer hardware education is rare. Yet research into remote methods of teaching can be useful not only in pandemics but also for improving educational opportunities in disadvantaged communities or during wartime~\cite{rajab2018theea}.

\citeauthor{astatke2012}~\cite{astatke2012} is one of the few to describe teaching embedded systems using remote hardware access. However, it relies on a specialized educational board and costly oscilloscopes. Furthermore, the custom board that was used is no longer supported and no obvious replacement exists.

A close relative to teaching embedded systems is teaching robotics, often also interacting with real hardware. Nevertheless, the reader has to keep in mind that robotics courses tend to focus on more high-level concepts and the requirements for contact with hardware are less strict~\cite{holowka2020, birk_2020}.

A platform providing remote access to robots was described by \citeauthor{kulich2013}~\cite{kulich2013}. The authors report students extensively used remote access outside of normal working hours or from home. Thus, contrary to previous years, most students could test their code on real hardware. With the deployment of the remote access system, the enrollment for the course increased. However, the system relies on a custom IDE plugin, web interface, robot, and the robot labyrinth – implying high development and maintenance costs. Furthermore, their approach is too custom-tailored to robotics to adapt to our use case.

Although the COVID-19 pandemic caused a surge of emergency remote teaching~\cite{emergency-remote-teaching} (a rapid switch to online education), comparative studies of online and offline education are limited. \citeauthor{chung2018} managed to do that to some extent~\cite{chung2018} -- they formulated a list of necessary components that online education of embedded systems requires to be viable: 1) student-teacher interaction, 2) independent learning skills, 3) well-designed learning content and 4) tangible support. However, they operate in the context of high schools, so their conclusions may not be directly applicable in our setting.

The work of \citeauthor{Brooks2021}~\cite{Brooks2021} compared the difficulties of a rapid transition to online education doing interviews across multiple lecturers from a selected university. From this, multiple patterns emerged: The workload was much higher, lectures were switched to asynchronous mode and engaging students was more difficult.

\section{Course Transformation}
\label{sec:course-redesign}

This section describes the context of conducted changes, possible options, and our solution in more detail.

\subsection{Context and Syllabus}

The course functions as an introduction to programming for microcontrollers. It is an optional part of the curriculum at the Faculty of Informatics, Masaryk University, Czechia. It is a medium-sized IT faculty with about 280 employees and 2\,300 students. The curriculum is mostly focused on software aspects and computer science theory, giving this course an important position as an introduction to low-level hardware. The course enrollment is usually about 30 students. We do not expect any prior hardware experience or electrical engineering knowledge. The only requirement is an understanding of the C language.

For most students, this is the first contact with microcontrollers and low-level hardware. We focus on peripherals and hardware-specific aspects, such as handling input noise and limited resolution. A good example is the proper handling of a button press (it can oscillate instead of producing a simple binary signal~\cite{BouncingButton}). This experience is rather difficult to simulate, as the range of potential problems is quite wide. Given how easy it is to give students access to real hardware, simulating such processes has not been developed much.

We use a standard development board FRDM-K66F~\cite{K66F}, designed by NXP. These boards contain a microprocessor and a set of peripherals (RGB LED, IMU, audio). We also provide students with additional external modules, such as USB-to-UART bridges, pressure sensors, joysticks, etc. We do this to give students at least basic skills of interconnecting the modules and the board. From now on, we use the term \textit{development kit} to denote the board alongside the set of extra modules.

The course starts with the introduction of basic concepts such as polling and interrupts for input readings. Then we continue with basic communication interfaces such as SPI and UART. In our examples, we are trying to show our students that even though we have only one core, they have to think of race conditions and different states their firmware may be in. These issues may not arise as much in simulations because the behavior of simulated hardware is usually more orderly and issues of real hardware may not arise.
At the end of the course, we introduce some higher-level interfaces such as USB and Ethernet.

In standard offline semesters, students are required to do weekly assignments on the topic introduced in the given week. The weekly assignments are simple and students are expected to spend no more than 30 minutes solving them. Students are allowed to complete the assignment either at the end of the class or after the class.

Apart from the weekly assignments, there is the semester project. The difficulty varies with regards to the specific assignment, but it takes around eight hours on average. As part of the project, students are required to present a short protocol regarding the project solution, describe the solution, and give arguments for the chosen method. For the project, students choose and set up the peripherals individually.

The course is pass/fail (i.e., not graded). To successfully pass the course, students must solve most weekly assignments and defend their final project explaining their decision-making.

\subsection{Available Teaching Setups}

\begin{table*}[t]
\caption{Comparison of four scenarios for teaching embedded systems at university level}
\label{tab:comparison}
\renewcommand\tabularxcolumn[1]{p{#1}}
\renewcommand{\arraystretch}{1.5}
\newcommand{\detail}{\newline\small\it}
\rowcolors{2}{white}{gray!20}
\begin{tabularx}{\textwidth}{>{\raggedleft\sc\arraybackslash}p{2cm}|Xp{3.6cm}XX}
& Scenario I\newline\textbf{Hardware at School} & Scenario II\newline\textbf{Remote Hardware Access} & Scenario III\newline\textbf{Hardware at Home} & Scenario IV\newline\textbf{Virtualized Hardware} \\ \hline
Remote learning
    & not supported
    & mostly OK \detail teachers need to be at the lab
    & OK \detail fully remote
    & OK \detail fully remote\\
Contact with hardware
    & direct \detail kits available at school
    & passive \detail teachers setup the HW
    & direct \detail kits available at home
    & virtual \detail only virtualized ideal version \\
Cost
    & medium \detail kit for each present student
    & medium \detail kit for each online workstation
    & high \detail kit for each student
    & low \detail if virtualized locally
    \\
Time Availability
    & limited \detail seminars and open lab hours 
    & anytime
    & anytime 
    & anytime \\
Software Availability
    & easy \detail local licensed installations
    & easy \detail local licensed installations
    & may be complicated \detail possible licensing problems
    & usually available \detail mostly proprietary\\
Complexity for Students
    & medium \detail students have to handle HW, \newline but assistance is simple
    & simple \detail HW is handled by teachers, \newline students \enquote{just use} it
    & high \detail students have to handle HW, \newline yet assistance is complicated
    & usually simple \detail interacting with simulator can be simpler than HW \\
Assisting Students
    & simple \detail interaction in the lab
    & medium \detail teachers can see HW in the lab, student can share screen
    & complicated \detail screen sharing, but HW at home and support only via camera
    & simple \detail screen sharing gets all info
    \\
Work with Peripherals
    & simple \detail handled by students, teacher \newline assistance easily available
    & medium \detail handled by teachers
    & complicated \detail handled by students, teacher \newline assistance is difficult
    & complicated \detail peripherals are often difficult to realistically simulate \\
Hardware Maintenance
    & students and teachers \detail students working with kits differently, re-assembly required
    & teachers \detail uniform setup by teachers
    & students \detail separate hardware setups, \newline problematic without assistance
    & non-existent \detail virtual environment can be easily reset
    \\
Serendipitous learning
    & high potential\detail people meeting in the lab
    & possible \detail people sharing hardware
    & low \detail people working individually
    & low \detail people working individually
    \\
Custom Projects
    & easy \detail access to all extra hardware
    & complicated \detail teacher must setup the projects
    & medium \detail reduced access to parts and tools
    & medium \detail limited and less fun
    \\
\end{tabularx}
\renewcommand{\arraystretch}{1}
\end{table*}

In retrospect, we see four alternative setups with regard to hardware interactions:
\begin{description}
    \item[Hardware at school.]
        Students visit the laboratory, work with local hardware and interact with teaching assistants in person.
    \item[Remote hardware access.]
        Hardware is present at the laboratory, and students connect to it remotely from anywhere.
    \item[Hardware at home.]
        University lends the hardware to the students who work with it independently at home.
    \item[Virtualized hardware.]
        There is no real hardware present. Students work with simulated hardware.
\end{description}
Each of these four setups has its advantages and disadvantages. A comprehensive comparison based on ten years of experience teaching hardware-based courses is summarized in \cref{tab:comparison}. The rest of this section elaborates on some of the more complex features.

Transitioning to teaching online can require a different amount of change in the course materials. Remote hardware access can reuse all the materials from the setup with hardware at school as only the remote connection layer is new. With hardware at home, the lack of teacher assistance demands more detailed  manuals for hardware interactions or adjusted seminars. Virtualization usually changes the tool stack and thus also requires new tutorials or even adjustments in the course of the seminars.

\begin{figure*}[t]
    \centering
    \includegraphics[width=\textwidth]{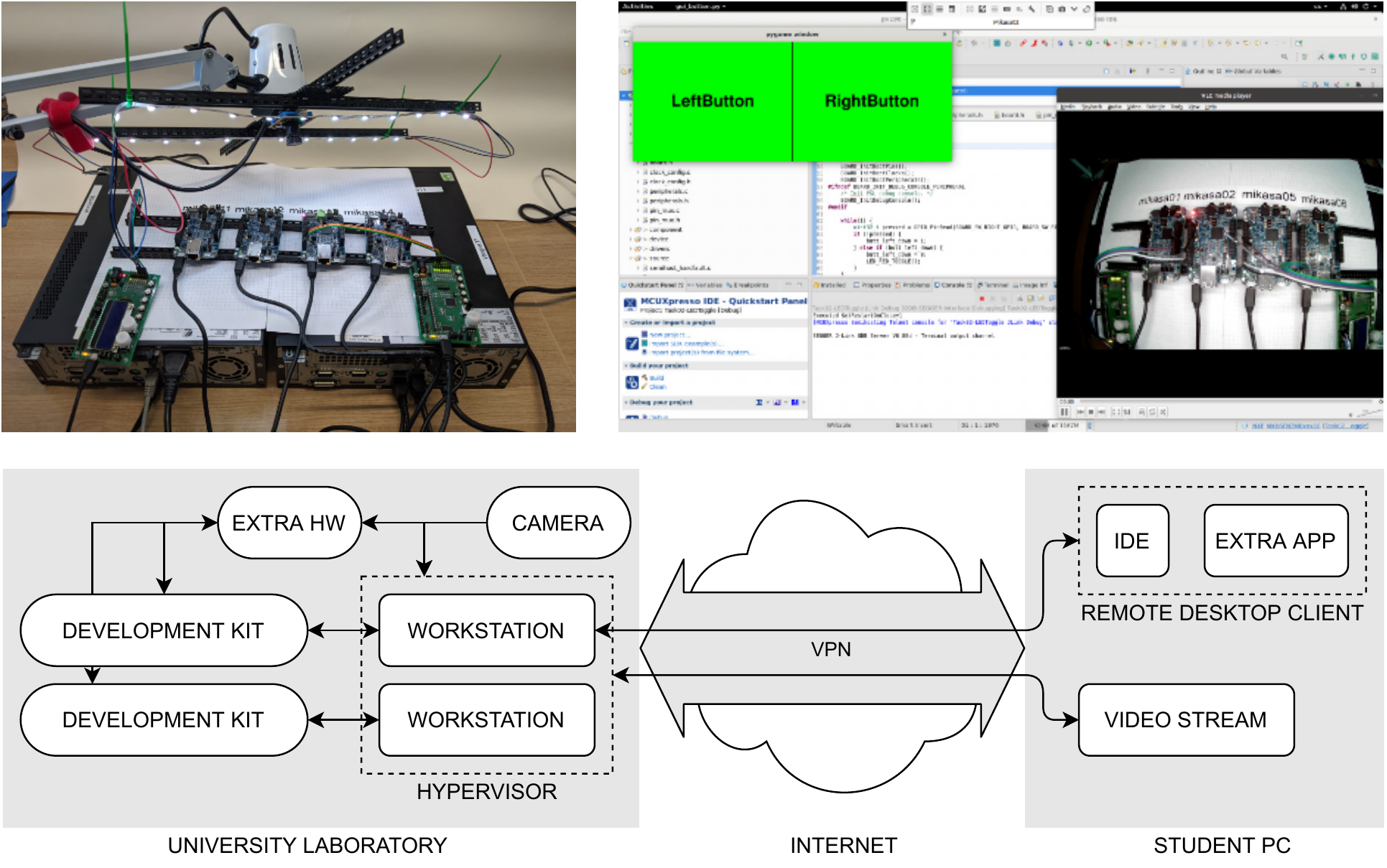}
    \caption{Setup overview of the remote hardware access approach. The photo in the top left shows the hardware setup of two hypervisors with four workstations and four development kits. The screenshot in the top right picture represents the student view of the system. The diagram on the bottom visualizes the technology stack used for working with hardware buttons.}
    \label{fig:overview}
\end{figure*}

Licensing can be a complicated issue. Historically, we used proprietary software that required a license server in the network. Recently, we moved to development kits without a need for a license server, which simplified the workflow. Licenses may be a complication with hardware at home -- although accessing the university license server may be possible, floating licenses and connections are often complicated. Licensing issues are also relevant for virtualization. Free virtualization software (e.g., Qemu emulator~\cite{qemu}) works well, but it is problematic to emulate peripherals. Alternatives exists but require a license~\cite{FVP} or focus on the Arduino platform~\cite{ATC}, which is not sophisticated enough for our needs.

When comparing to teaching robotics, there is a significant difference of focus: Robotics courses tend to focus on high-level algorithms and control, while courses on embedded systems often concentrate on the low-level details of hardware control~\cite{holowka2020,birk_2020}. For example, it is reasonable in robotics that students will understand algorithms like A* from programming a virtual mouse in a virtual labyrinth. The same students will not understand the complexities of sensing a color from a sensor unless they experience all the hardware limitations and challenges. This makes virtualization in robotics more feasible, resulting in plenty of available robotic simulators useful for the education.

The topic of cost is also very variable. Lending hardware to students scales badly with the number of students, making it an interesting option only for courses with a low number of students. On the other hand, a standard offline approach with one seminar at a time requires development kits only for that seminar (and some in reserve). The virtualization option can also be quite expensive: Apart from the potential license fees, powerful virtualization hardware may be needed.

\subsection{Our Transition to Remote Teaching}

We designed our solution with multiple design goals. Firstly, it should be as simple and modular as possible, ideally avoiding any new development of custom software systems. Secondly, it should reuse existing infrastructure and utilize existing stable technologies, preferably open ones. If successful, this would result in a low-maintenance solution enabling fast transition (in a matter of weeks from design to testing). Furthermore, a standard modular technology stack would ensure discontinued tools can be easily switched for others.

An overview of our setup, based on the remote hardware access setup, can be seen in \cref{fig:overview}. We have prepared a cluster of hypervisors, each hosting two virtual machines. Each virtual machine was one workstation to which students connected remotely. Each workstation had a pre-installed IDE and was connected to one development kit. To see the kit behavior, a camera stream was provided per four kits. The whole infrastructure was accessible for students remotely at any time.

In this section, we describe key aspects of the setup. It should work as a useful inspiration for anyone attempting to set up a similar setup. The detailed automated setup in the form of the Ansible playbook (an open-source system configuration tool~\cite{ansible}) is available at \href{https://github.com/koniarik/teaching-embedded-remotely}{github.com/koniarik/teaching-embedded-remotely}.

\paragraph{Course Structure and Communication}

Although one of our goals was to keep the content unchanged, the structure had to be adjusted a bit. We introduced the concept of demos to the existing seminars: We recorded all necessary information for each week and gave the recording to the students.
The content of seminars was the same as the content of demos, just done differently and interactively.

We used a dedicated Discord server for all communication~\cite{discord}, as it provides both the ability to hold group voice calls and recorded asynchronous instant messaging. Furthermore, our students were already familiar with the service from their personal lives -- we assumed this would make it really simple for them to contact us.

\paragraph{Assessment}

Because the lacking student access to hardware was complicating semester projects, we decided to adjust the assessment and drop the project entirely. Instead, we increased the difficulty of weekly assignments to take approximately 90 minutes to solve (compared to the 30 minutes previously). While increasing the difficulty, it had been decided to evaluate assignments more strictly. Even with that, 29 out of 34 students solved all assignments correctly, while the remaining five were asked to do additional work. 

\paragraph{Hypervisors and Workstations}

The infrastructure was based on virtualization -- it consisted of base hypervisor hardware and virtualized workstations. For the hypervisors, we decided to repurpose older lab computers that were previously used to work with the hardware. Apart from handling the virtual machines, each hypervisor had a camera connected. This was used to stream a video of development kits from up to four workstations (see \cref{fig:overview}).

Each hypervisor handled two virtual workstations. Each workstation was a standalone development environment for one person. We connected one development kit to each workstation. Apart from these, we connected some additional hardware needed to work with the boards (prime example are extra boards in \cref{fig:overview} used to simulate button interactions).

\paragraph{Networking}

The laboratory has an isolated network governed by the local server. We connected hypervisors to this network and enabled port forwarding from selected high ports of the server to the virtual workstations. To secure the access we utilized the existing university VPN service, filtering at the firewall for selected ports. In summary (\cref{fig:overview}), we allowed only remote desktop and SSH access to the workstations (programming the development kits) and to the hypervisors (video streams, generating sensor inputs). The network setup was fully automated using Ansible.

\paragraph{Generating Inputs}

To let students test their solutions, sensors on the development kits needed controlled inputs (e.g., pressing buttons or moving the kit to engage the accelerometer). To generate such inputs, we employed extra development kits connected to the hypervisors. These controlled the motors periodically tilting the student kits to generate accelerator data or generated digital signals simulating button presses. Unfortunately, we found no tools to enable comfortable remote control, but creating a minimalist graphical interface accessible was more than sufficient (for example, see the button controls in \cref{fig:overview}).

\paragraph{IDE}

To minimize the necessary changes in the course, we wanted to stick to the free IDE from the manufacturer (MCU Expresso~\cite{NXPIDE} by NXP). In our experience, it has the best hardware support and beginner-friendly tools for configuration and debugging. This is crucial for the introductory course, as we do not want to overwhelm the students. For example, the configuration of peripherals is much simpler with graphical tools (like the one present in MCU Expresso) than just by code.

There were two installation options built on existing infrastructure: 1) Students installing IDEs on their computers and using a remote server (e.g., GDB) for programming and debugging or 2) students connecting by remote desktop to the workstations and using the IDE within the remote desktop environment. The remote GDB server appeared to be an unfeasible option with the selected IDE, as there were multiple connection issues. Furthermore, the remote desktop environment had the benefit of an easier setup.

\paragraph{User Management}

Because the virtual machines were shared and the resources were limited, we needed a solution to schedule and manage workstation access. Aiming for simplicity, we used a simple shared Google Spreadsheet. For our case (a course with about 30 students) a shared table with the honor system turned out as perfectly sufficient. Nevertheless, larger classes may require a dedicated reservation system or access enforcement.

\section{Evaluation}
\label{sec:evaluation}

In this section, we retrospectively evaluate our fast transformation to remote teaching from both student and teacher perspectives.

\subsection{Student Perspective}

At the end of the course, the students were asked to complete a short questionnaire summarizing their views and feelings. The questions asked can be categorized into three groups: previous experience, issues, and ideas for the future. It was answered only by 11 out of 34 students, which may introduce a self-selection bias.

Most of the students had no problem with the technical content of the course (even though it was the first contact with hardware programming for many of them). The largest issue reported (almost unanimously) concerned the development environment. The remote desktop access caused (sometimes huge) latency in responses to user inputs. As one of the students reported: \textit{\enquote{The idea of remote boards with webcams is interesting. Although the implementation could be better, mainly the poor performance of workstations [was problematic].}}

On the one hand, the largest advantage students had seen in the remote setup was the possibility to access hardware literally anytime (day and night, workdays and weekends). In a standard teaching setup, students can access the hardware laboratory only on workdays from 8:00 am to 8:00 pm. Students had also seen a great advantage in the pre-recorded seminars. This allowed them to re-watch sections that they found difficult or revisit topics covered weeks ago. On the other hand, there was a group of students who would prefer the standard offline version: 
\textit{\enquote{Yes, I would prefer to visit the lab. I think it would be much easier to directly work with the hardware and visiting the lab is not really a problem.}}

Looking at these opinions, we wonder about the possibility of having both at the same time. This way, students could have the benefits of anytime access from home and yet receive full interactive education at the laboratory during the seminar.

\subsection{Teacher Perspective}

The strongest observations of teachers concerned the effects of communication (asynchronous, textual, recorded), the necessity to maintain the hardware, and different homework assessments.

The overall adjustments in communication did not pose any significant issues. With the (pre-recorded) demo videos and study materials, most of the students could work independently. The presence of recorded interactions had a strong effect: Compared to usual (recordless) face-to-face interactions or voice calls, pre-recorded videos, instant messaging, and emails are recorded, allowing students to get back to them when necessary. In our setup, the difference was noticeable: Firstly, students knew much more from the lecture material and did not want to have it repeated during the seminars. Secondly, we had a much lower frequency of repeated questions, as students got into a habit of searching the history of our communication platform first. On both levels, the availability of records was beneficial.

With the development kits not under students' direct control, the occasional necessary hardware restarts had to be performed by teachers (many things can go wrong in hardware and preventing them all was not realistic). This led to the need for at least one teacher to be present in the lab throughout the week. Furthermore, the teachers had to set up new hardware each week (e.g., new sensors or different interactions). Connecting all peripheries and verifying the setup for all kits was a tedious and time-consuming process. It was not the best long-term solution, but it was manageable in our small case.

\begin{table}[t]
    \centering
    \caption{Student's self-reported time spent on the course, compared across the standard and remote semester}
    \label{tab:semester-stats}
    \begin{tabular}{>{\raggedleft\sc\arraybackslash}m{2.7cm}|ll }
        & \textbf{Standard} & \textbf{Remote    }\\
        \hline
        Max   & 80 h    &125 h\\
        Min&   35 h  & 20 h \\
        Median &60 h & 50 h\\
        Average    &63.3 h & 57.8 h\\
        \hline
        Student responses &12 out of 24 & 19 out of 34
    \end{tabular}
\end{table}

Even though the homework assignments were considerably more difficult than in the previous semester, the assessment was still similarly simple (pull student code, run test cases, evaluate). In most cases, the submitted solutions received the full score. Overall, it seems we managed to retain the course difficulty level comparable to the standard in-person semester (see \cref{tab:semester-stats}).

\section{Lessons Learned}
\label{sec:lessons-learned}

After comparing the options and describing our approach in the previous sections, we draw lessons learned from our experience.
\begin{description}[itemsep=0.5em,leftmargin=2em]
    \item[Remote access approach works.]
        The atypical approach to remote teaching of embedded hardware proved to work well enough. Although there is definitely room for improvement, it enabled us to keep at least indirect contact with real hardware.
    \item[Complex custom-made systems are not necessary.] 
        We were able to assemble the system out of existing stable components without the need for custom development. That increases robustness and simplifies maintenance.
    \item[Demos are useful.]
        We saw an unexpected improvement in students' ability to understand what was given to them (decreasing the usual necessity to repeat information multiple times). The students themselves appreciated the recorded demos and the fact that they could revisit older topics.
    \item[Anytime access is appreciated.]
        Based on the feedback from students, the most appreciated aspect was the ability to work anytime (expectable, as having to visit the laboratory to continue the work on homework or project can be inconvenient). A similar conclusion was also reported by \citeauthor{kulich2013}~\cite{kulich2013}.
    \item[Hybrid access is preferred.]
        To our surprise, students would prefer to have both in-person and remote access to hardware to enjoy the benefits of the lab and the possibility of working from home when necessary (e.g., homework deadline).
    \item[Changes to the configuration are tiresome.]
        In the remote hardware access setup, any change to the hardware configuration is tiresome to do, as the teachers have to adjust the setup on all kits. Although manageable in our case, we consider effort put into minimizing this need as worthwhile.
    \item[Remote desktop has latency.]
        As reported by students, the remote desktop access has poor performance – although that is possibly affected by the quality of students' connection. Nevertheless, we should always expect there will be some students with a poor connection.
    \item[Familiar and less formal communication platform.]
        We saw increased usage of the Discord communication platform outside of seminars compared to the traditional ways of contacting teachers (email, university information system). It is not clear whether this was due to the specifics of this semester or because of the different nature of the technology, but we assume that we can clarify this in the next offline semester.
    \item[Teachers should do retrospection.]
        Deeply analyzing the course and performing adjustments in retrospect lead us to a better and much clearer understanding of what we did and what we want to do in the future. Writing this experience report pressed us to verbalize our thoughts and feelings and deeply compare and contrast the possible teaching approaches.
\end{description}
On the one hand, the Discord communication platform and pre-recorded demos are features we want to transfer also into the traditional in-person semesters. Based on our understanding of why they work, keeping them should preserve the benefits even for in-person teaching.

On the other hand, we are still hesitant to retain the possibility of  remote hardware access in parallel to the in-person access for the upcoming semester. Further work would be required to stabilize the current solution and space/maintenance requirements are also non-trivial. Nevertheless, the system is a clear choice for us should remote teaching be necessary again.

\section{Conclusion}
\label{sec:conclusion}

Due to a pandemic, we had to transition a university course on embedded hardware programming to enable remote education. We compared four principally different options. In the end, in-lab teaching was unavailable due to the pandemic, lending hardware to students was too costly and to emulating hardware avoided the intended hardware interactions. We decided to try an atypical solution of remotely accessed hardware.

Although remote hardware access has been used in teaching before~\cite{chung2018,kulich2013,astatke2012}, all previous works featured custom-built tools with costly development and high necessary maintenance. We investigated the viability of avoiding complex custom-made systems. Our experience shows remote hardware access is a feasible alternative, deployable in a matter of weeks using existing open technologies.

\bibliographystyle{ACM-Reference-Format}
\bibliography{main}

\end{document}